\documentclass[11pt,letterpaper]{article}
 \pdfoutput=1
\usepackage{jcappub}

\usepackage{caption}
\usepackage{floatrow}
\usepackage{journals}

\usepackage{feynmp}

\usepackage{subfig}

\usepackage{bm}

\def\be{\begin{eqnarray}}
\def\ee{\end{eqnarray}}
\def\kk{\bm{\Bbbk}}

\def\k{\bm{k}}

\def\rec{\mathrm{rec}}

\def\L{\mathcal{L}}

\def\x{\bm{x}}
\def\q{\bm{q}}

\def\G{\bm{G}}

\def\s{\bm{s}}

\def\d{\delta_D}

\def\g{\bm{g}}

\def\st{\mathrm{st}}
\def\SS{\bm{d}}

\title{Towards an Optimal Reconstruction of Baryon Oscillations}
\author{Svetlin Tassev$^{a}$ and Matias Zaldarriaga$^{b}$}
\affiliation{ \sl $^{a}$ Center for Astrophysics, Harvard University, Cambridge, MA 02138, USA\\
 \sl $^{b}$ School of Natural Sciences, Institute for Advanced Study, Olden Lane, Princeton, \\NJ 08540, USA
}

\abstract{ The Baryon Acoustic Oscillations (BAO) in the large-scale structure of the universe leave a distinct peak in the two-point correlation function of the matter distribution. That acoustic peak is smeared and shifted by bulk flows and non-linear evolution. However, it has been shown that it is still possible to sharpen the peak and remove its shift by undoing the effects of the bulk flows. We
 propose an improvement to the standard acoustic peak reconstruction. Contrary to the standard approach, the new scheme has no free parameters, treats the large-scale modes consistently, and uses optimal filters to extract the BAO information. At redshift of zero, the reconstructed linear matter power spectrum leads to a markedly improved sharpening of the reconstructed acoustic peak compared to standard reconstruction. 
} 
\notoc
\begin{document}
\maketitle

%\documentclass[11pt,letterpaper]{article}
%\pdfoutput=1
%\usepackage{epstopdf}
%\usepackage{jmy}
%\bibliographystyle{JHEP}

%\usepackage{bm}

\section{Introduction}\label{intro}

The Baryon Acoustic Oscillations (BAO) are an important target for studies of large-scale structure (LSS) in the universe. They appear as an acoustic peak in the 2-point correlation function of the galaxy distribution  \cite{2005ApJ...633..560E}. The scale of the acoustic peak corresponds to the distance sound waves have traveled in the photon-baryon fluid until recombination. 
 That scale ($\approx 150\,$Mpc) is well in the linear regime, and bulk matter flows and non-linear physics can affect the scale only at a percent level \cite{2007ApJ...664..660E}. Yet, these effects introduce a $\sim10\,$Mpc broadening of the acoustic peak, which in turn degrades the accuracy in the determination of the acoustic scale. Most of this smearing is due to large-scale flows \cite{tassev}, whose effects can be reversed at least partially  \cite{2009PhRvD..79f3523P,2010ApJ...720.1650S,2012arXiv1202.0090P} in a process known as reconstruction. 
As LSS surveys become larger and more complete, increasingly more advanced algorithms to compensate for the degradation of the acoustic peak will be needed to fully utilize the observations. Ongoing and future surveys of LSS are expected to achieve a sub-percent level accuracy in the determination of the acoustic scale, making the development of such algorithms the more urgent.
 
In this paper we propose a new scheme for the reconstruction of the acoustic peak, the goal of which is to recover the linear density field starting from the non-linear matter density field.   The method relies on a recent model \cite{2012JCAP...12..011T} for the non-linear density field, where one approximates the trajectories of Cold Dark Matter (CDM) particles as given by a scale- and time-dependent rescaling of the particle displacements in the Zel'dovich approximation (ZA) \cite{zeldovich}. 
 That model allows for the construction of optimal filters, with which to iteratively process the matter density field, so as to compensate for the smearing and shifting of the acoustic peak induced by the large-scale matter flows. This is done without introducing any free parameters, unlike the standard reconstruction scheme which relies on finding an optimal smoothing scale by trial-and-error.  
 
 An important ingredient of the method described in this paper is its treating of the large-scale modes. Standard reconstruction recovers them in a Zel'dovich-like approximation, which implies (as we will see below) that one can choose to recover well either the large-scale modes ($k\lesssim 0.2h/$Mpc for redshift of $z=0$); or the short-scale modes ($k\gtrsim 0.2h/$Mpc for $z=0$) by degrading the large-scale reconstruction. Our method ameliorates  this trade-off considerably. This results in a significantly improved sharpening of the acoustic peak in the matter correlation function.

In this paper we apply our reconstruction scheme only on the matter distribution in real space, as we rely on previous results \cite{2012JCAP...12..011T} (see also \cite{tassev}) to model the CDM matter density field. Once those results are extended to cover biased tracers, it should be straightforward to write down the proposed reconstruction scheme for biased tracers of the underlying density field. However, it is not yet clear how much the reconstruction will be improved in that case.

In Section~\ref{sec:st} we give a review of standard reconstruction, while in Section~\ref{sec:OV} we give a very brief overview of our method. Then  in Section~\ref{sec:new} we derive a \textit{pseudo}-optimal\footnote{By ``pseudo-optimal'' we mean that our derivation would have produced an optimal estimator if the fields involved were Gaussian. However, the density field in the (mildly) non-linear regime is non-Gaussian. Therefore, all we can say for sure is that our method provides better reconstruction results than the standard reconstruction scheme -- a result which we obtained only after actually applying the method to N-body simulations.} reconstruction algorithm, given the model of \cite{2012JCAP...12..011T}. We find that that algorithm converges slowly, and therefore in Section~\ref{sec:IC} we present an efficient algorithm which provides a good initial guess for the pseudo-optimal algorithm in Section~\ref{sec:new}. We compare the proposed reconstruction scheme with standard reconstruction in Section~\ref{sec:results}, and then we summarize our results in Section~\ref{sec:summary}. In the Appendices we give the details behind our numerical implementation of the proposed reconstruction scheme.

\section{Standard reconstruction}\label{sec:st}
Let us start with a review of standard reconstruction \cite{2009PhRvD..79f3523P,2010ApJ...720.1650S,2012arXiv1202.0090P}, which will allow us to set up some notation.\footnote{In this paper we compare only against the one-step ZA procedure described in \cite{2010ApJ...720.1650S}. Those authors investigate the effect of incorporating second-order Lagrangian perturbation theory in their reconstruction scheme, as well as performing several iterations of their method. The improvements they find are only marginal, and therefore we do not focus on those modifications here. } The exact position, $\x_p$, of particle $p$ in Eulerian space  (obtained e.g. from N-body simulations) is given by
\be\label{eom}
\x_p(\q,\eta)=\q_p+\s(\q_p,\eta)\ ,
\ee
where $\q_p$ is the Lagrangian particle position; $\eta$ is conformal time; and $\s$ is a stochastic vector field. As an example, at linear order in the overdensity we have 
\be\label{ZA}
\partial_{\q}\cdot \s_z(\q,\eta)=-\delta_L(\q,\eta)=-\delta_L(\q,\eta_0)D(\eta) \ ,
\ee
where $\delta_L$ is the linear density field, $D$ is the growth factor such that $D(\eta_0)=1$. The above equation corresponds to the Zeldovich approximation (ZA) \cite{zeldovich}, hence the subscript $z$.

The overdensity field, $\delta[\x_p](\x)$, corresponding to the distribution of particles $\x_p$, is a functional of the particle positions and is proportional to the sum over the number of particles falling in a volume element around $\x$.  As an illustration, starting from a set of CDM particles sitting at coordinates $\x_p$ given in Eulerian space ($\x$), one could easily obtain the overdensity ($\delta$) using for example a Cloud-in-Cell (CiC) assignment. 
In the continuum limit, this particle representation results in
\be\label{delta}
\delta[\q_p+\s(\q_p,\eta)](\x)=\int d^3 q \d\left(\x-\q-\s(\q,\eta)\right)-1\ ,
\ee
where $\delta_D$ is the Dirac delta function.

%
%Reconstruction then can be treated as an inversion problem, where one starts with the density field above, and tries to obtain the displacement field, $\s(\q,\eta)$. Then one tries to evolve the displacement field backward in time to the linear regime, where its divergence gives the linear overdensity:
%\be
%\delta(\x,\eta) \to \s(\q,\eta) \to \s_z(\q,\eta) \to %\delta_L(\q,\eta)
%\ee

Standard reconstruction then works as follows. The density field, $\delta$, is Gaussian smoothed (with a filter $W_G=\exp(-\Sigma^2 k^2/2)$), and then a large-scale  displacement field, $\s_R(\x)$, is obtained to linear order from eq.~(\ref{ZA}):
\be\label{sR}
\s_R(\k)=i\frac{\bm{k}}{k^2} W_G(k)\,\delta(\k) \ ,
\ee
where $R$ stands for ``Reconstructed'', and the wavevectors above correspond to Eulerian Fourier space.
One then displaces the particle positions, (\ref{eom}), by $-\s_R$, obtaining the following density field:
\be\label{deltaS}
\delta_s\equiv\delta[\x_p-\s_R(\x_p)]\ ,
\ee
which to leading order reduces to $(1-W_G)\delta_L$.
Note that $\x_p-\s_R(\x_p)$ is an approximation to the Lagrangian positions of the particles, since we have subtracted the (dominant) large-scale displacements. Therefore, $\delta_s$ is really an approximation to the short-scale part of the linear overdensity in Lagrangian space, $\delta_L(\q)$.

To get the large-scale modes in the standard reconstruction method, one moves a uniform random particle field, $\q_u$ ($u$ running over the random particles), by $\s_R$ to obtain
\be\label{deltaLambda}
\delta_\Lambda\equiv \delta [\q_u+\s_R(\q_u)]\ ,
\ee
which to leading order reduces to $W_G\delta_L$.
The resulting approximation, $\delta_L^\st$, to the linear density is then given by 
\be\label{addSL}
\delta_L^\st=\delta_s+\delta_\Lambda\ ,
\ee
where ``$\st$'' stands for standard reconstruction.\footnote{In the actual standard reconstruction, the plus in both (\ref{deltaLambda}) and (\ref{addSL}) is replaced with a minus. This is because in the standard reconstruction scheme one displaces the random points by $-\s_R$ to get $\delta_\Lambda$.}

Note that eq.~(\ref{deltaLambda}) effectively mixes Lagrangian and Eulerian coordinates. This is because $\q_u$ can be treated as particle positions in Lagrangian coordinates, since $\delta[\q_u]$ is uniform by construction. Yet, $\s_R$ is an Eulerian vector field, which we evaluate at the Lagrangian instead of the Eulerian positions. This difference of course is irrelevant at leading order, but we are interested in the higher-order corrections. Indeed, if we cared only about the linear order, we could have simply written $\delta_L=\delta$, which can hardly be called a reconstruction. 

Even if $\s_R$ was the correct Lagrangian space displacement field, what eq.~(\ref{deltaLambda}) would give us is the density field in the ZA corresponding to the displacements $\s_R$. Therefore, the cross-correlation coefficient between $\delta_L$ and $\delta_\Lambda$ exhibits a Gaussian decay on a scale approximately given by the large-scale rms particle displacements. One finds the same behavior for the cross-correlation coefficient between $\delta_L$ and $\delta$ (see e.g. \cite{tassev}). It is the presence of $W_G$ that makes the former decay smaller by reducing the rms particle displacements, which is why standard reconstruction works at all.

\section{Brief overview of our method}\label{sec:OV}

From our discussion of standard reconstruction so far, we can see that there are two main places where standard reconstruction can be improved: 1) One can try to obtain optimal filters for processing the non-linear density field, instead of relying on filters chosen by hand ($W_G$); 2) One can treat the large-scale modes consistently by not mixing Eulerian and Lagrangian coordinates, and by obtaining the large-scale density field as minus the divergence of the Zel'dovich displacements, instead of by using the Zel'dovich-like approximation of (\ref{deltaLambda}). These issues are solved in our reconstruction scheme outlined below.

Before we proceed with the description of our method, one should note that there are two main approaches to reconstruction: ``forward'' and ``backward''. In a forward reconstruction, we try to match the non-linear $\delta$ with a $\delta[\s_z]$ corresponding to our reconstructed $\s_z$ -- a fitting which we perform in Eulerian space. In a backward reconstruction, we try to move around the data particles, $\x_p$, responsible for the non-linear $\delta$ in such a way so as to get as uniform a field, $\delta[\x_p-\s_z]$, as possible. In other words, in a backward reconstruction we try to match the uniform initial conditions, and therefore this kind of fitting can be thought of as being performed in Lagrangian space.

One should note that if one performs backward reconstruction on real galaxy catalogs, one implicitly  assumes that the data particles (the galaxies) cover Lagrangian space uniformly -- a highly non-trivial assumption for biased tracers. So, a forward scheme  should be the scheme of choice for real-life data. Yet, in experimenting with different algorithms, we found that backward schemes are generally both more stable and faster converging.

Therefore, the reconstruction scheme proposed in this paper consists of two stages. In the first stage we use a well-motivated but yet ad hoc backward scheme for efficiently constructing a good guess for $\s_z$. That guess is then used as an initialization for our forward quasi-optimal second stage of the reconstruction.

In the first stage (Section~\ref{sec:IC}) we solve iteratively for a Lagrangian displacement field $\bm{d}(\q)$ which moves the data particles (e.g. galaxies) to form a uniform distribution (see (\ref{consistency0})).
We then relate the displacement field $\bm{d}$ to the Zel'dovich displacement, $\s_z$, using a linear model (see (\ref{corrD})). Therefore, the large-scale $\s_z$ is given by a Wiener-filtered $\bm{d}$, while the large-scale reconstructed linear $\delta$ is given by minus the divergence of the large-scale $\s_z$. This is in sharp contrast with standard reconstruction, where the large-scale modes are reconstructed in a Zel'dovich-like approximation (we will return to this point later on). The Wiener filtering, however, loses information about the short scales, which we then recover in a manner analogous to standard reconstruction, (\ref{deltaS}).

The resulting reconstructed $\s_z$ (given in (\ref{sznew})) is then used as an initial guess for the quasi-optimal forward second stage (Section~\ref{sec:new}) of our reconstruction scheme. In that stage, we try to maximize the likelihood function given by (\ref{lnL}), which uses information about the  statistics of the initial density field, and tries to match the non-linear $\delta$ to an LPT-inspired model for the non-linear density field. The result of that procedure is an $\s_z'$ (see (\ref{iter})), which is very well-correlated to the true Zel'dovich displacement field. Yet, $\s_z'$ is a non-linear Wiener-like filtered displacement field, and thus it is suppressed at short scales. We recover them by filtering $\s_z'$ with an inverse Gaussian (see (\ref{finalFix})) to obtain $\hat \s_z$.  Our final result for the reconstructed linear density  field is then given by minus the divergence of $\hat\s_z$, according to (\ref{ZA}).

In the next two sections we will describe in detail the two stages of our reconstruction scheme. First we start our discussion by deriving the quasi-optimal second stage of our method. The second stage is independent on the first stage of the reconstruction, which only supplies an initialization for the second stage. Then, in Section~\ref{sec:IC} we give the detailed recipe for our choice of initialization, constituting the first stage of the method.

\section{Improving standard reconstruction}\label{sec:new}

In this section, we will derive a pseudo-optimal iterative scheme for performing the BAO reconstruction. We will start by writing down a model for the CDM density field, presented first in \cite{2012JCAP...12..011T}. Then, we will write down the corresponding Likelihood function, $\mathcal{L}$, and from there we will obtain an iterative algorithm, with which we will solve for $\s_z$, which maximizes $\L$. That iterative algorithm will be our reconstruction scheme, because knowing $\s_z$ is equivalent to knowing $\delta_L$ according to (\ref{ZA}).

\subsection{Modeling the density field in the mildly non-linear regime}

To make progress we use the following split for the non-linear density field proposed in \cite{2012JCAP...12..011T}:
\be\label{master}
\delta(\k)=R_\delta (k)\delta_1(\k)+\delta_{MC}(\k)\ , \ \ \mathrm{with } \ \ \delta_1\equiv \delta[\x_p=\q_p+( R_z*\s_z)(\q_p)]\ ,
\ee
where $\delta_{MC}$ is a mode-coupling term, such that $\langle \delta_1(\k)\delta_{MC}^*(\k)\rangle=0$. Here $\k$ is in Eulerian Fourier space and the $*$ denotes convolution in Lagrangian space. We introduced $R_\delta (k)$ and $R_z (\Bbbk)$ -- the so-called density and displacement transfer functions, which are given in \cite{2012JCAP...12..011T}. Here $\k$ denotes a wavevector corresponding to Eulerian Fourier space, while $\kk$ denotes a wavevector in Lagrangian Fourier space. These transfer functions are given by a linear fit to the density and the displacement fields respectively:
\be
R_\delta(k)=\frac{\langle \delta(\k) \delta_1^*(\k)\rangle}{\langle | \delta_1(\k)|^2\rangle} \ ,\ \ 
 R_z(\Bbbk)=\frac{\langle \s(\kk)\cdot \s_z^*(\kk)\rangle}{\langle | \s_z(\kk)|^2\rangle}\ .
\ee
These transfer functions can be cheaply obtained by calibration with simulations because of their small sample variance. We refer the reader to \cite{tassev} and \cite{2012JCAP...12..011T} for further discussion on this point, as well as to \cite{2012JCAP...12..011T} for the details of the above model for $\delta$ used in this paper.

\subsection{The Likelihood Function}

The likelihood function, $\L$, for $\s_z$ given the model above is:
\be\label{lnL}
-\log \L \supset \int d^3k \frac{|\delta(\k)-R_\delta(k)\delta_1(\k)|^2}{P_{MC}(k)} +\int d^3\Bbbk \,\Bbbk^2\frac{|\bm{s}_z(\kk)|^2}{P_L(\Bbbk)}\ ,
\ee
where we made the difference between Lagrangian and Eulerian wavevectors explicit. The power spectrum $P_{MC}$ is the power spectrum of $\delta_{MC}$, but can include other noise terms, such as Poisson noise if $\delta$ is sampled with discrete tracers. In writing $\L$ above, we made the simplifying assumption that $\delta_{MC}$ is Gaussian.

The pseudo-optimal\footnote{We have assumed a Gaussian $\delta_{MC}$, which makes our calculation no longer optimal.} $\s_z$ for the Likelihood function above can be found by taking the functional derivative of $\log \L$ with respect to $\s_z$ and setting to zero. For that we need the following derivative
\be
\frac{\partial \delta_1^*(\k)}{\partial s^*_{z,j}(\bm{\Bbbk})}=\int \frac{d^3q}{(2\pi)^3}e^{i\k\cdot\big(\q+(R_z*\s_z)(\q)\big)}ik_je^{-i\bm{\Bbbk}\cdot \q}R_z({\Bbbk})\ ,
\ee
where we denote functional derivatives by $\partial$ so that we can distinguish them from the overdensity. Here again the Lagrangian wavevector is denoted with  $\bm{\Bbbk}$, while the Eulerian -- by $\k$. To derive the above equation we wrote down $\delta_1(\k)$ by Fourier transforming (\ref{delta}) with $\s\to R_Z*\s_z$. Thus, for a quantity $A(\k)$ we can write:
\be\label{Z}
\int d^3k \frac{\partial \delta_1^*(\k)}{\partial \s^*_{z}(\bm{\Bbbk})} A(\k)&=&\int \frac{d^3q d^3k}{(2\pi)^3}A(\k) e^{i\k\cdot\big(\q+(R_z*\s_z)(\q)\big)} i\k e^{-i\bm{\Bbbk}\cdot \q}R_z(\bm{\Bbbk})\nonumber\\
&=&\int \frac{d^3q}{(2\pi)^3}\, d^3x\d\big(\x-\q-(R_z*\s_z)(\q)\big)\nabla_{\x}A(\x)R_z({\Bbbk})e^{-i\bm{\Bbbk}\cdot \q}\ .
\ee
Combining (\ref{Z}) and (\ref{lnL}) we obtain that the maximum of the likelihood function is at $\s_z$ that satisfies: 
\be
0&=&-\frac{1}{2}\frac{\partial\log \L}{\partial \s^*_{z}(\bm{\Bbbk})}=\\\nonumber
&=&\frac{\Bbbk^2}{P_L(\Bbbk)}\s_z(\bm{\Bbbk})-R_z({\Bbbk})\int \frac{d^3q}{(2\pi)^3}e^{-i\bm{\Bbbk}\cdot\q}\bm{\nabla}_{\x}\bigg[\int d^3k\, e^{i\k\cdot\x}\,\frac{R_\delta(k)\big(\delta(\k)-R_\delta(k)\delta_1(\k)\big)}{P_{MC}(k)}\bigg]\ ,
\ee
where $\x$ is evaluated at $\bm{x}=\q+(R_z*\s_z)(\q)$, and one should note that $\delta_1$ is a functional of $\s_z$.
The integral in $\q$ above gives the Lagrangian Fourier transform of the Eulerian gradient of the quantity in the square brackets. Thus, we can write the above expression as
\be\label{derL}
\Bbbk^2\s_z(\bm{\Bbbk})=R_z({\Bbbk})P_L(\Bbbk)
\mathcal{F}_{\kk\q}\mathcal{F}^{-1}_{\x\k}\left[i\k
\frac{R_\delta(k)\big(\delta(\k)-R_\delta(k)\delta_1(\k)\big)}{P_{MC}(k)}
\right]\ ,
\ee
where $\mathcal{F}_{\kk\q}$ denotes the Fourier transform from $\q$ to $\kk$, while $\mathcal{F}^{-1}_{\x\k}$ denotes the inverse Fourier transform from $\k$ to $\x$. The reason $\mathcal{F}_{\kk\q}\mathcal{F}^{-1}_{\x\k}$  does not equal 1, is because between those two transforms we map the coordinates according to $\bm{x}=\q+(R_z*\s_z)(\q)$.
At first order, however, the difference between $\x$ and $\q$ (and between $\k$ and $\kk$) is irrelevant and $\delta_1=-\bm{\nabla}\cdot(R_z*\s_z)$,  and thus we obtain:
\be\label{derL1}
-i\k\cdot \s_z(\k)=\frac{(R_zR_\delta)^2P_L}{P_{MC}+(R_zR_\delta)^2 P_L}\times \frac{\delta(\k)}{R_zR_\delta}\ ,
\ee
which is the standard Wiener filter result for a linear model relating the displacement and the non-linear overdensity. Thus, we should think about the result for $\s_z$ in (\ref{derL}) as coming from a non-linear Wiener-like filtering of $\delta$. Note, however, that in what follows we use the exact result given by (\ref{derL}), and not the approximation given by (\ref{derL1}).

\subsection{Maximizing the Likelihood function iteratively}
We would like to find the maximum of the Likelihood function iteratively using the usual Newton-Raphson method. Thus, we write the updated value $\s_z'$ as:
\be
s_{z,i}'(\bm{\Bbbk})=s_{z,i}(\bm{\Bbbk})-\left[\frac{1}{\delta_D(\bm{0})}
\left\langle\frac{\partial^2\log\L}{\partial \s_{z}(\bm{\Bbbk})\partial \s_{z}^*(\bm{\Bbbk})}\right\rangle\right]^{-1}_{ij}\frac{\partial\log\L}{\partial s_{z,j}^*(\bm{\Bbbk})}\ ,
\ee
where we used homogeneity (which is why we have the factor of $1/\delta_D(\bm{0})=(2\pi)^3/V$, where $\bm{0}$ is the zero wavevector, and $V$ is the (Lagrangian) volume of the survey/N-body simulation); and as usual we used the Fisher matrix instead of the Hessian of the Likelihood function in order to simplify the inversion.

Next we need take the second derivative of the likelihood function in analogy with the first derivative. We find:
\be
&&-\frac{1}{2\delta_D(\bm{0})}\frac{\partial^2\log\L}{\partial s_{z,i}(\bm{\Bbbk})\partial s_{z,j}^*(\bm{\Bbbk})} = \delta_{ij}\frac{\Bbbk^2}{P_L(\Bbbk)}+\\\nonumber
&&+\frac{R_z^2(\Bbbk)}{\delta_D(\bm{0})}\int d^3kd^3qd^3\tilde q (2\pi)^{-6} \frac{R^2_\delta(k)}{P_{MC}(k)}k_ik_j e^{i\k\cdot\big(\q+(R_z*s_z)(\q)-\tilde{\q}-(R_z*s_z)(\tilde{\q})\big)
}e^{-i\kk\cdot \big(\q-\tilde{\q}\big)}\nonumber\\
&&+\frac{R_z^2(\Bbbk)}{\delta_D(\bm{0})}\int d^3kd^3q(2\pi)^{-3}\frac{R_\delta(k)\left(
\delta(\k)-R_\delta(k)\delta_1(\k)\right)}{P_{MC}(k)}
e^{i\k\cdot\big({\q}+(R_z*s_z)(\q)\big)
}k_ik_j\nonumber\ .
\ee
The integral over $\q$ in the last term above reduces to $\delta_1^*(\k)$. Therefore, using eq.~(\ref{master}) and the fact that by construction $\langle\delta_1^*\delta_{MC}\rangle=0$, we find that the ensemble averaged value of the last term is zero. Ensemble averaging the rest of the terms, we find that the Fisher matrix is given by
\be
\left\langle -\frac{1}{2\delta_D(\bm{0})}\frac{\partial^2\log\L}{\partial s_{z,i}(\bm{\Bbbk})\partial s_{z,j}^*(\bm{\Bbbk})}\right\rangle
=\delta_{ij}\left\{
\frac{\Bbbk^2}{P_L(\Bbbk)}+\frac{R^2_z(\Bbbk)}{3}
\left\langle\mathcal{F}_{\kk\q}\mathcal{F}^{-1}_{\x\k}\left[
k^2
\frac{R^2_\delta(k)}{P_{MC}(k)}\right]\right\rangle
\right\}\ ,
\ee
with $\x=\q+(R_z*\s_z)(\q+\tilde\q)-(R_z*\s_z)(\tilde\q)$ for an arbitrary $\tilde\q$. For the second term above, we used both homogeneity and isotropy to simplify the ensemble average. The former brings out a Dirac delta function which cancels $1/\delta_D(\bm{0})$ (and allows us to use an arbitrary $\tilde\q$), while the latter tells us that the result must be proportional to $\delta_{ij}$. 

Combining the results above, we end up with the following iterative solution for $\s_z$:
\be\label{iter}
\s_{z}'(\bm{\Bbbk})&=&\s_{z}(\bm{\Bbbk})-W(\Bbbk)
\left[\s_z(\bm{\Bbbk})-\frac{R_z({\Bbbk})P_L(\Bbbk)}{\Bbbk^2}
\mathcal{F}_{\kk\q}\mathcal{F}^{-1}_{\x\k}\left[i\k
\frac{R_\delta(k)\big(\delta(\k)-R_\delta(k)\delta_1(\k)\big)}{P_{MC}(k)}
\right]\right],\nonumber\\
W^{-1}(\Bbbk)&\equiv&
1+\frac{R_z^2(\Bbbk)P_L(\Bbbk)}{3\Bbbk^2}
\left\langle\mathcal{F}_{\kk\q}\mathcal{F}^{-1}_{\tilde \x\k}\left[
k^2
\frac{R^2_\delta(k)}{P_{MC}(k)}\right]\right\rangle\ ,
\ee
where $\tilde\x=\q+(R_z*\s_z)(\q+\tilde\q)-(R_z*\s_z)(\tilde\q)$, $\x=\q+(R_z*\s_z)(\q)$, and $\delta_1=\delta[\q+(R_z*\s_z)(\q)]$ as before. The resulting expression above is written in a way suitable for implementation into a numerical code, since all integrals have been replaced by Fourier transforms.\footnote{Note that the ensemble average in the Fisher matrix can be easily performed analytically, in the same way one calculates the analytical Zel'dovich power spectrum, $P_Z$. However, unlike the $P_Z $, the final Lagrangian space integrals in the Fisher matrix are not as trivial. So, we stick with an N-body approach  to map $\k\to\kk$ through $\mathcal{F}_{\kk\q}\mathcal{F}^{-1}_{\tilde \x\k}$. One should note that whether or not one approaches the Fisher matrix by using N-body simulations, writing a code to do that mapping is still required for the first derivative of the Likelihood function.} To simplify our notation, we revert to denoting both Eulerian and Lagrangian wavevectors with $\k$, unless stated otherwise.

\begin{figure}
  \centering
  $z=0$\hspace{0.5\textwidth} $z=1$
  \\
  \subfloat{\includegraphics[width=0.5\textwidth]{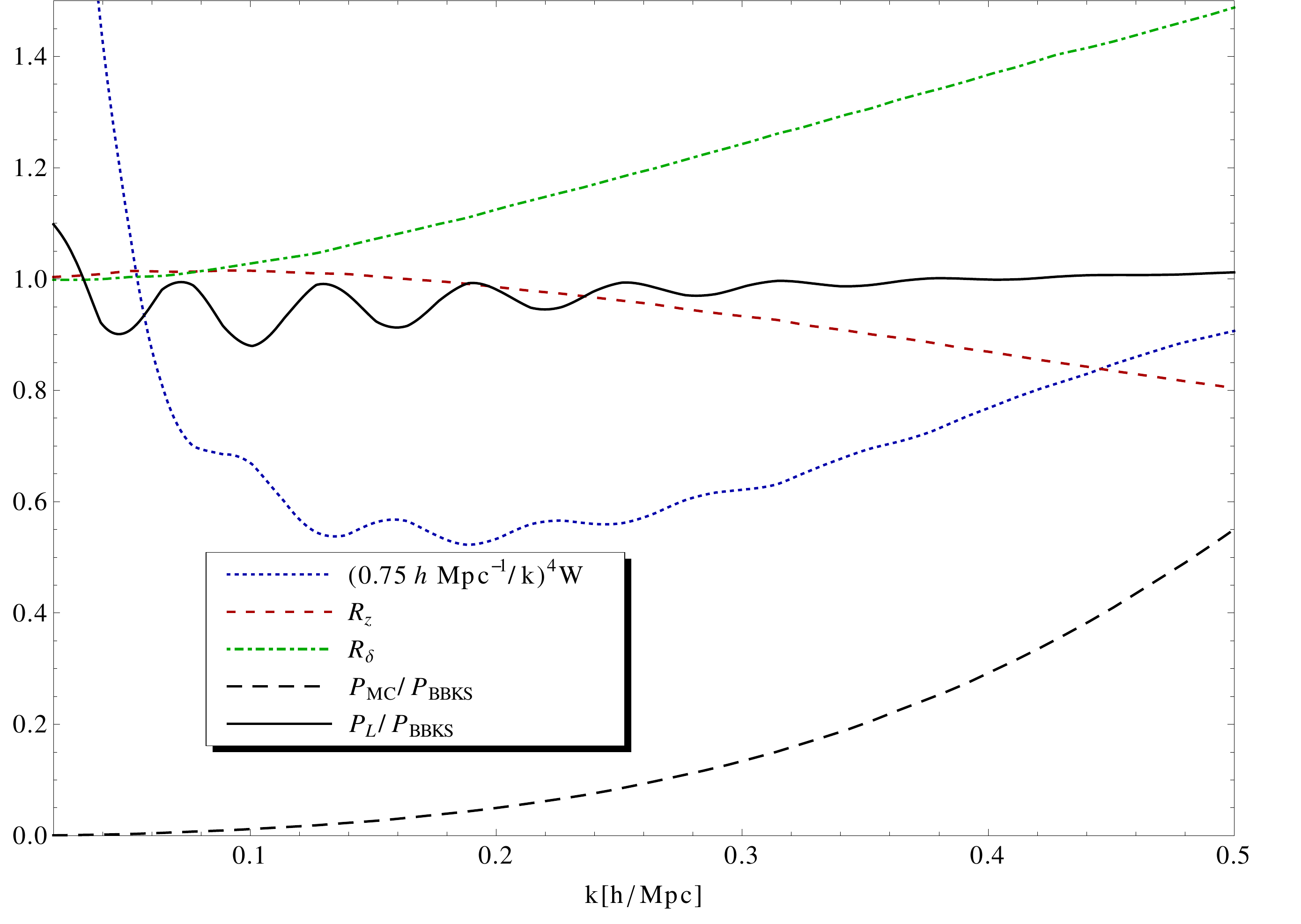}}                
  \subfloat{\includegraphics[width=0.5\textwidth]{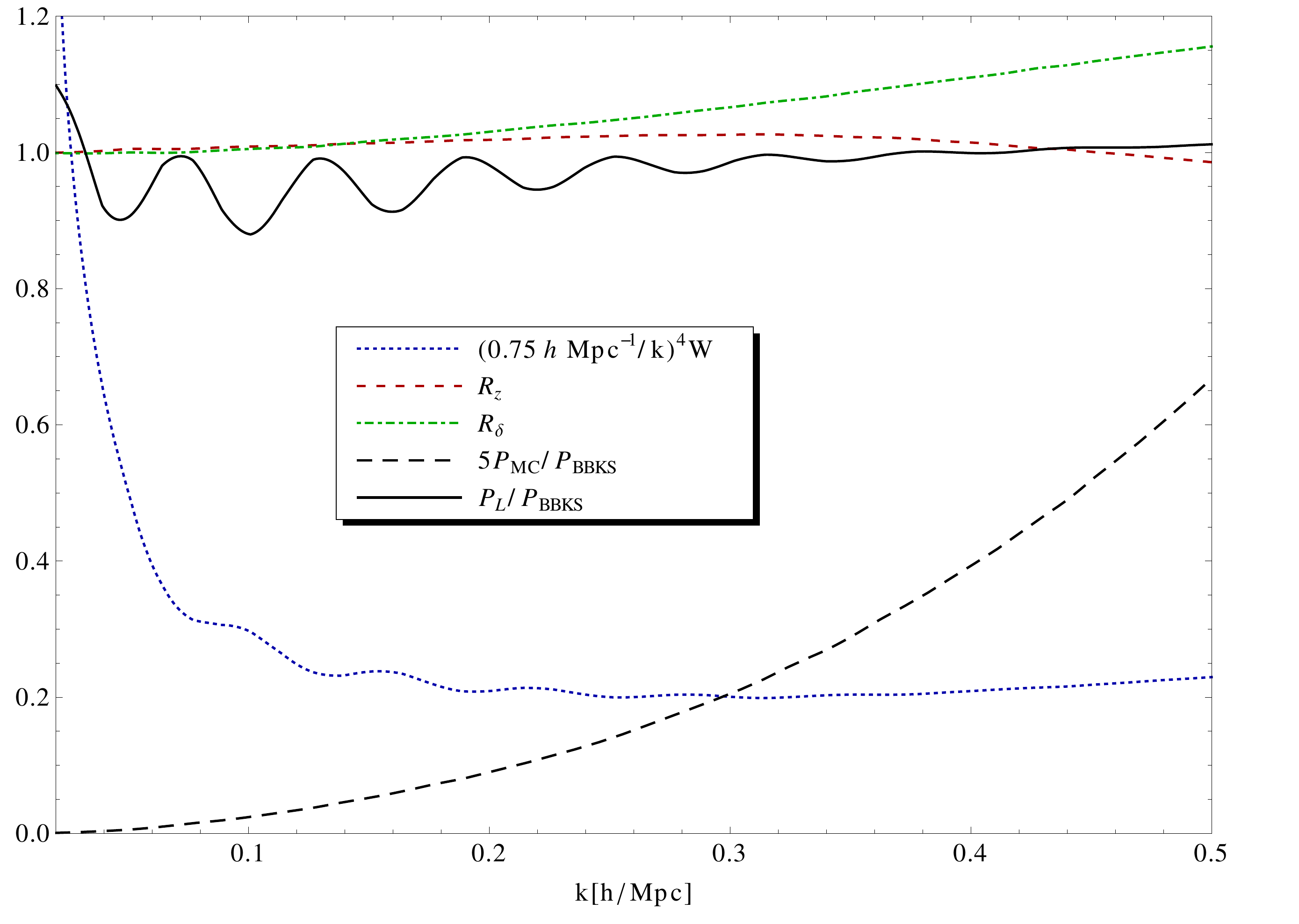}}
  \caption{We show the different quantities entering in the iterative solution for $\s_z$, eq.~(\ref{iter}). The power spectra are divided by a smooth BBKS \cite{BBKS} power spectrum with shape parameter $\Gamma=0.15$ in order to highlight the wiggles due to the BAO; and $P_{MC}$ is multiplied by 5 for $z=1$ to fit in the plot. Note that we have rescaled the inverse of the Fisher matrix $W$, to highlight the scale dependence.  }
  \label{fig:Filters}
\end{figure}

In Figure~\ref{fig:Filters} we show the different quantities entering in eq.~(\ref{iter}) for reference. In the actual numerical reconstruction we smooth out all wiggles in these quantities, so that wiggles are not introduced by hand, and only appear if present in the data. Note that the inverse Fisher matrix, $W$, is strongly weighted towards high $k$, which compensates the small-$k$ weighting coming from the $P_{MC}$ in the denominator in (\ref{iter}). The Fisher matrix is obtained by performing the average in (\ref{iter}) over all N-body boxes described in Section~\ref{sec:results}, where for each box we have averaged over 100 randomly chosen $\tilde \q$'s.

The transfer functions $R_z$ and $R_\delta$ are 1 for small $k$, and deviate from 1 only for $k\gtrsim k_{NL}$ (see \cite{2012JCAP...12..011T} for further discussion). The mode-coupling power, $P_{MC}$, vanishes at small $k$, which is a direct result of the fact that $\delta_1$ and $\delta$ are highly correlated for those scales \cite{2012JCAP...12..011T}. 

\subsection{Discussion}

The result of this section is equation (\ref{iter}), which gives a pseudo-optimal iterative solution for the Zel'dovich displacement field in Lagrangian space, given a non-linear density field in real space. There are, however, several important caveats to this result, which we discuss below.

Note that (\ref{iter}) involves highly non-linear functionals of $\s_z$. Thus, one is not guaranteed that that iterative algorithm will find the global maximum of the Likelihood function, and not some local maximum. For example, the density field $\delta_1$ is insensitive to a swapping of pairs of particles. Such a swapping is in principle penalized by the term proportional to $|\s_z|^2$ in $\mathcal{L}$, (\ref{lnL}), but it is not outright forbidden. Moreover, in the numerical calculation we do not want to introduce BAO wiggles by hand. Therefore, as mentioned before, we smooth out all wiggles from the quantities in (\ref{iter}), most notably -- $P_L$ and $W$. This automatically makes the algorithm not-completely-optimal (because of the modified $W$), and it does not converge to a true maximum of $\mathcal{L}$ (because of the modified $P_L$, entering in the first derivative of $\mathcal{L}$). These same problems are further exacerbated because of our assumption of a Gaussian $\delta_{MC}$ in writing down the Likelihood function. Even if the complications above were not present, one is always left with the problem of restoring the high-$k$ reconstructed power (see below), which has been killed by the Wiener-like filtering of the algorithm above.

The above discussion implies that we should better start the algorithm, (\ref{iter}), as close to the true global maximum of $\mathcal{L}$ as possible. This is especially true, since we find the method to be very slowly converging. So, in Section~\ref{sec:IC} we construct a non-optimal (not even pseudo-optimal) but very efficient algorithm to provide us with a good initial guess for $\s_z$, which can then be fed into (\ref{iter}). With that guess for $\s_z$, we find that the algorithm above requires only $\mathcal{O}(3)$ iterations to converge (see Appendix~\ref{sec:details} for further details).

Note that the discussion so far assumes that we are actually calculating the linear power spectrum  directly from $\s_z$ using (\ref{ZA}). However, in the standard calculation of optimal quadratic estimators for Gaussian random fields, none of the problems above arise. Why is this the case? The reason is that for Gaussian fields one would integrate the likelihood function (\ref{lnL}) over all reconstructed realizations, $\s_z$, and then optimize with respect to $P_L$, and not with respect to $\s_z$ as we did above. That would result in a $P_L$, which is not affected by the smoothing applied by the Wiener-like filtering (e.g. (\ref{derL1})), because the Wiener filter drops out, as well as any possible bias introduced by a wrong initial guess for $P_L$, the $R$'s, etc. However, integrating (\ref{lnL}) with respect to $\s_z$ is an immensely complicated task. Therefore, we have to make some approximation. And the approximation we do is finding $P_L$ from the pseudo-optimal $s_z$ using (\ref{ZA}). It is important, however, first to correct $\s_z$  as much as possible at least for the smoothing introduced by the Wiener-like filtering  (e.g. (\ref{derL1})). This is what we do next in an ad hoc manner.

The discussion above implies that the residual between the obtained $\s_z'$ and the true $\s_z$ may still contain recoverable information about the BAO. 
In order to recover it, we write the relation between the true $\s_z$ and the recovered $\s_z'$ as:
\be
\s_z=R_z'\s_z'+\s_{MC}'\ ,
\ee
with $\langle\s_z'\s_{MC}'\rangle=0$. A linear fit gives $R_z'=\langle s_{z,i} s_{z,i}'^*\rangle/\langle|s_{z}
'|^2\rangle$. 
%Therefore we can see that  $$k^2|\s_{MC}'|^2\equiv \d(\bm{0})P_{MC}'=\d(\bm{0}) P_L(1-\rho'^2)\ ,$$ where $\rho'$ is the cross-correlation between $\s_z$ and $\s_z'$. Note that $\rho'$ contains wiggles of negligible amplitude (see the thick solid line in Fig.~\ref{fig:cross}), which therefore cannot compensate the wiggles in $P_L$. Thus, as long $\rho'$ is not 1, $\s_{MC}'$ should indeed contain important BAO information. 
We find that indeed both $R_z'$ and $\s_{MC}'$ contain important residual BAO information. 
So, we cannot use the trick of \cite{tassev} to find the transfer function and the mode-coupling piece from other realizations (or slightly wrong cosmologies). Otherwise, we would be putting wiggles in our results by hand. Instead, we should recognize that if the cross-correlation between $\s_z$ and $\s_z'$ is very close to 1, this means that we have captured the complex phases of the perturbations correctly; and all that is left is to fix their amplitudes. Thus, we write our final result for the reconstructed $\s_z$ as:
\be\label{finalFix}
\hat \s_z(\kk)=A_f(\Bbbk) \s_z'(\kk)\ ,
\ee
where $A_f$ is an (inverse) Gaussian fit to $A\equiv \sqrt{P_L/P_L'}$, with $\d(\bm{0})P_L'\equiv |\s_z'|^2\Bbbk^2$. We use a Gaussian fit to $A$, since we recognize that $A$ again contains wiggles, which we do not want to introduce by hand. When we perform the numerical calculations, we find that $\log( A_f)=(\Bbbk/(1.19h/$Mpc$))^2$ for $z=0$ and  $\log( A_f)=(\Bbbk/(0.97h/$Mpc$))^2$ for $z=1$, which are the best fit Gaussians for the scales relevant for the BAO: $\Bbbk<0.4h/$Mpc. Note that $A_f(\Bbbk)$ acts as the inverse of the Wiener-like filtering discussed above, thus restoring the high-$\Bbbk$ power in $\s_z$. In Section~\ref{sec:results}, we show results for the density field obtained by taking minus the divergence of $\hat \s_z$, which is obtained by combining (\ref{iter}) and (\ref{finalFix}). But before we show those results, let us present the algorithm we use for initializing (\ref{iter}).

\section{Constructing a guess displacement field by modifying the standard reconstruction scheme}\label{sec:IC}

We find that the pseudo-optimal algorithm presented in the previous section converges very slowly -- i.e. it is pseudo-optimal but is not efficient. Thus, we have to construct an efficient (although not even pseudo-optimal) algorithm to provide a guess displacement field for the pseudo-optimal algorithm of the previous section. Thus, the scheme presented in this section should be treated as providing a first-stage reconstruction, the results of which are then improved further by the second-stage reconstruction of the previous section. Inevitably, the scheme of this section is ad hoc, since we are only guided by the principle of choosing a scheme that is efficiently converging to a displacement field, which is well correlated with the true Zel'dovich displacement field. At the same time, we tried to make each of the steps of the algorithm below as reasonable as possible.

Therefore, we will utilize the fact that there are several places where the standard reconstruction scheme can be improved. First, it would be good to write down the large-scale linear density field not through the Zel'dovich-like approximation given by eq.~(\ref{sR}, \ref{deltaLambda}), but by using a more self-consistent approach to extract the large-scale linear density field. The other place where it can be improved is if we find a (more) optimal prescription for obtaining the filter $W_G$, which in the standard reconstruction is set to be a Gaussian with a smoothing scale given approximately by the rms particle displacements ($\sim 15\,$Mpc$/h$ at $z=0$).

\subsection{Obtaining a displacement field capturing the power in the non-linear density}

As we discussed in Section~\ref{sec:OV}, the first stage reconstruction of this section is a backward scheme. In the spirit of a backward scheme, let us first find a displacement field $\bm{d}(\q_p)$, which nulls the power in the density, and therefore results in:
\be\label{consistency0}
\delta[\x_p-\bm{d}(\q_p)]\approx 0 \ ,
\ee
where again $\x_p$ and $\q_p$ are the positions of the data particles in Eulerian and Lagrangian space, respectively.  We will solve for the above equation iteratively below in a way analogous to \cite{2010ApJ...720.1650S}. The two main differences between our method and the iterative extension to standard reconstruction proposed in \cite{2010ApJ...720.1650S} is that: 1) we derive optimal filters (under a set of assumptions) with which we process the data instead of relying on filters chosen by hand; 2) we treat the large-scale modes consistently by not mixing Eulerian and Lagrangian coordinates.

\begin{figure}
  \centering
  $z=0$\hspace{0.5\textwidth} $z=1$
  \\
  \subfloat{\includegraphics[width=0.5\textwidth]{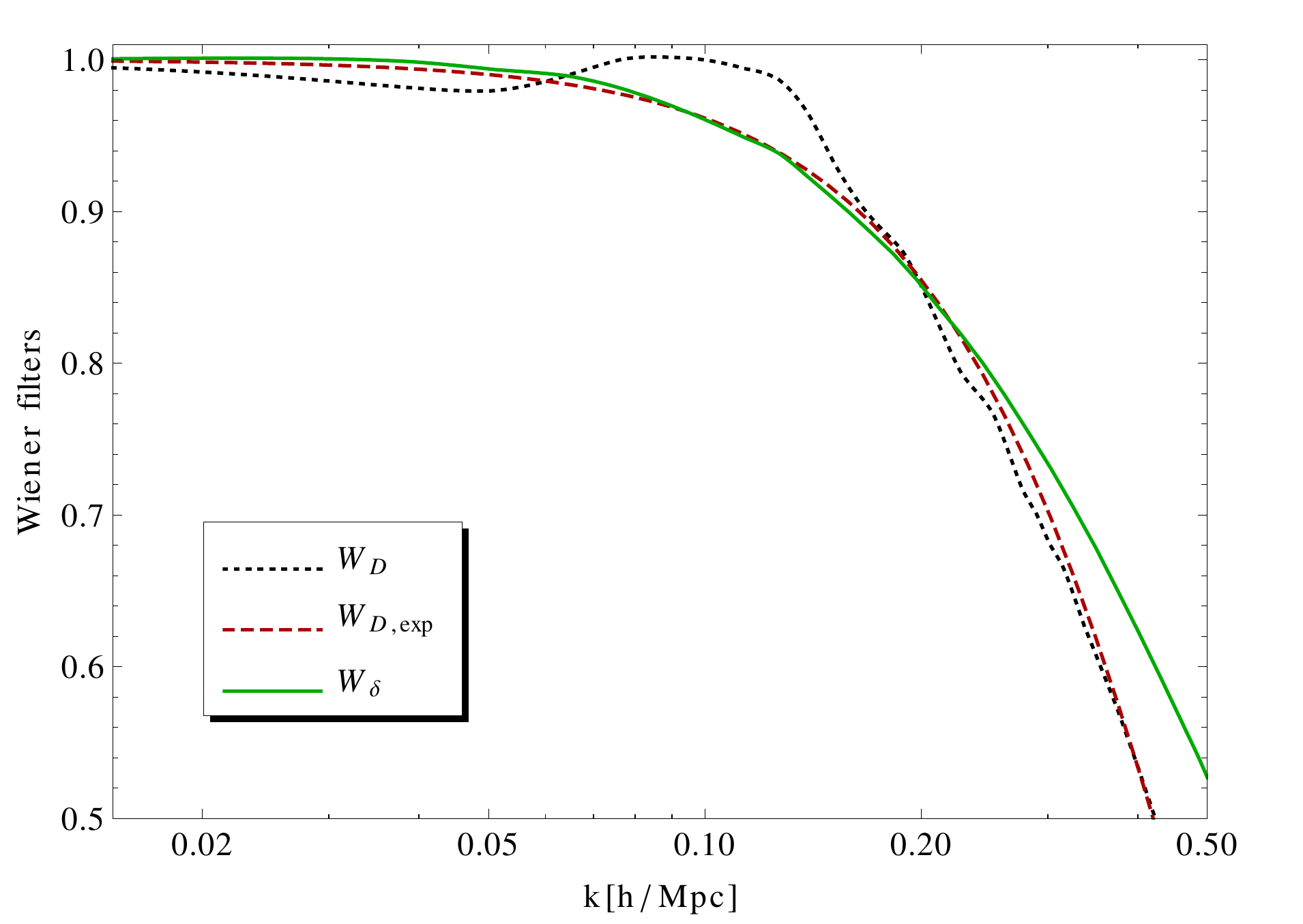}}                
  \subfloat{\includegraphics[width=0.5\textwidth]{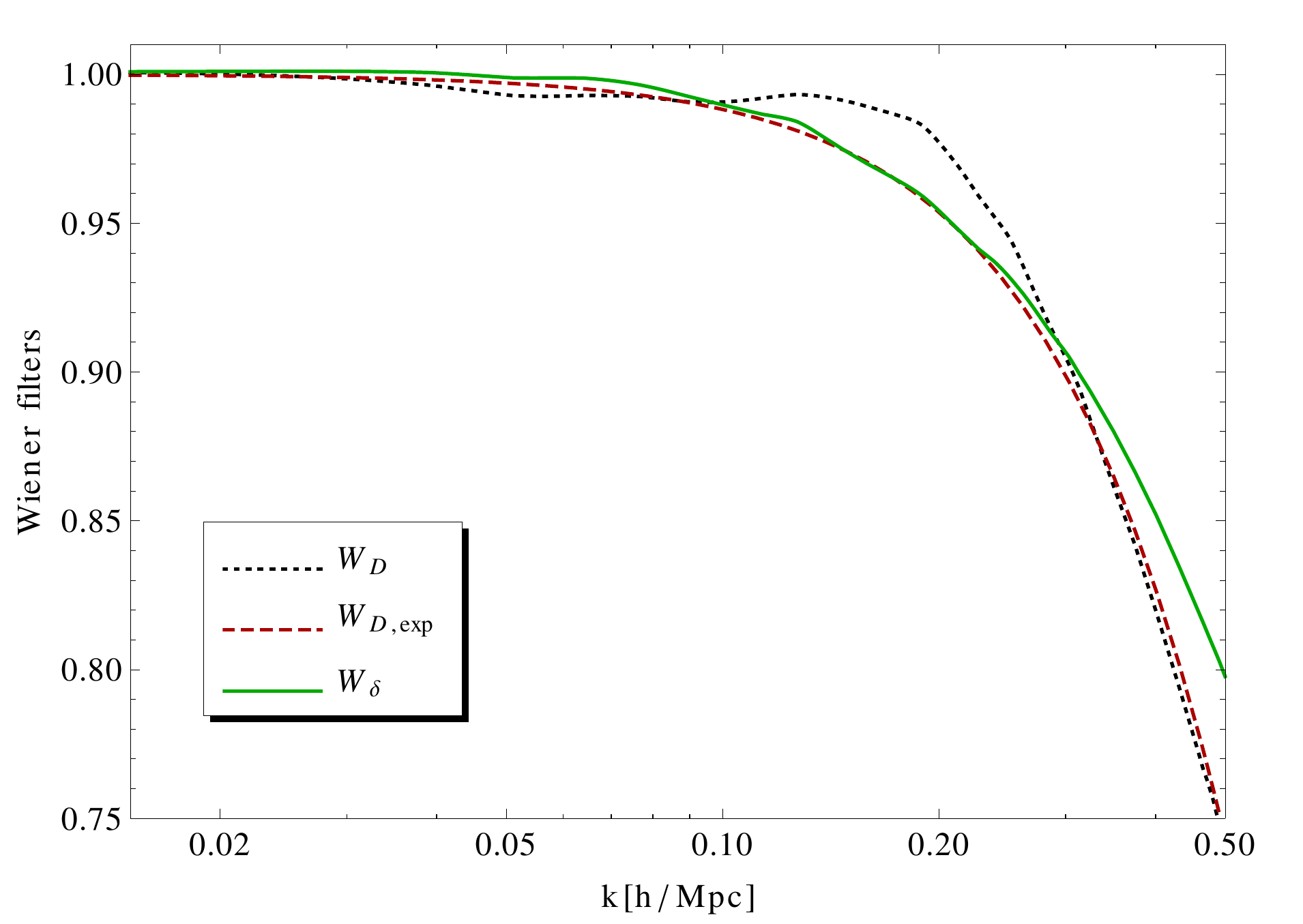}}
  \caption{The density and displacement Wiener filters discussed in the text. The filter, $W_D$, is calculated for $N=M=5$. In order to remove the BAO wiggles from $W_D$, in our reconstruction scheme we use a Gaussian fit to $W_D$, denoted by $W_{D,\mathrm{exp}}$. Note that $\k$ corresponds to Eulerian Fourier space for $W_\delta$, and to Lagrangian Fourier space for $W_D$ and $W_{D,\mathrm{exp}}$.}
  \label{fig:Wiener}
\end{figure}

We know that most of the recoverable information for $\s_z$ lies in $\delta_1$ according to the model (\ref{master}), and not in $\delta$.  Therefore as a first guess, we would like to match $\delta_1$:
\be\label{consistency}
\delta_1\approx \delta[\q_p+\bm{d}(\q_p)] \ \ \hbox{as a first guess, }
\ee
which requires obtaining $\delta_1$ from $\delta$. We do that by noticing that the model  (\ref{master}) is linear. Thus, assuming that $\delta_{MC}$ is Gaussian, the optimal solution is to use a Wiener filter to obtain a recovered $\delta_1$:
\be\label{delta1}
\delta_1^\rec(\k)=W_\delta(k)\delta(\k) \ , \ \ \mathrm{with}\ \ W_\delta(k)=\frac{\langle \delta(\k) \delta_1^*(\k)\rangle}{\langle | \delta(\k)|^2\rangle}\ ,
\ee
where $\k$ is in Eulerian Fourier space. 
This Wiener filter can be cheaply obtained by calibration with simulations because of its small sample variance, in analogy with the transfer functions $R_\delta$ and $R_z$ discussed above.
We plot the filter $W_\delta$ in Fig.~\ref{fig:Wiener} for redshifts $z=0$ and $1$. Note that the filter decays at about the non-linear scale (given by $k_{NL}=0.25h/$Mpc for $z=0$ and $k_{NL}=0.74h/$Mpc for $z=1$) as one can expect. It also features slight wiggles corresponding to the BAO wiggles. In order not to introduce those wiggles by hand, we smooth them out by fitting a B-spline function to $W_\delta$, and use that result in our numerical calculations.

To obtain $\bm{d}(\q_p)$, we apply the iterative method described in \cite{2010ApJ...720.1650S}, smoothing intermediate results for the density with $W_\delta$ (see Appendix~\ref{app:iter} for further details). In this iterative scheme we start with the true particle positions, which correspond to $\delta$. We smooth $\delta$ with $W_\delta$ to obtain a reconstructed $\delta_1$, and then derive the displacement field corresponding to it (using (\ref{ZA})). That  displacement field approximately matches $\bm{d}$ in (\ref{consistency}). We then displace the particles by minus that displacement field; recompute the density; smooth it with $W_\delta$; derive a new displacement field; and move the particles again. We iterate this procedure $N$ times, where fiducially $N=5$. After these 5 iterations, the power spectrum of the density field corresponding to the newly obtained particle positions ($\x_p^{(N)}$) is only about 10\% (for $z=0$) of the power in $\delta_L$ at $k\sim 0.75h/$Mpc and is strongly suppressed at lower $k$. Thus, the newly obtained particle positions, $\x_p^{(N)}$, result in a more or less uniform density field (up to $k\sim 0.75h/$Mpc). So, $\x_p^{(N)}$ should be close to the Lagrangian position of the particle $p$ apart from some short-scale corrections, which we will quantify later on (see eq.~(\ref{corrD})). 

That implies that for the purposes of the acoustic peak reconstruction, we satisfy eq.~(\ref{consistency0}), where $\SS(\q_p)$ is given by $\SS_p^{(N)}\equiv \x_p-\x_p^{(N)}$, which is the total displacement of particle $p$ after $N$ iterations. Therefore, the pairs $(\x_p^{(N)},\SS_p^{(N)})$ equal $(\q_p,\bm{d}(\q_p))$ up to some short-scale corrections. So, all we have to do next to obtain $\bm{d}(\q)$ is to interpolate those pairs to a grid. This can be done in variety of ways. Our method for accomplishing that is given in Appendix~\ref{app:interp}. It is an iterative method (the fiducial number of iterations ($M$) we choose for the method is $M=5$), which produces $d(\q)$ on a grid in Lagrangian space.

\subsection{Extracting the Zel'dovich displacement field}

Note that in the steps above we obtained some $\bm{d}$ satisfying (\ref{consistency0}), starting from a guess satisfying (\ref{consistency}). So, next we need to relate $\s_z$ to $\bm{d}$, compensating for any short-scale corrections, and non-linearities. We can always write that relation as follows (in Lagrangian Fourier space):
\be\label{corrD}
\bm{d}(\kk)=R_D(\Bbbk)\bm{s}_z(\kk)+\bm{s}_{MC}(\kk)\ ,
\ee
with $\langle s_{z,i}s_{MC,j}\rangle=0$. That allows for a linear fit for the new transfer function, $R_D$, giving $R_D=\langle \s_z\cdot \bm{d}\rangle/\langle |\s_z|^2\rangle$, which can be calibrated after applying the algorithm above to simulations. The equation above in principle should capture any problems arising from the inversion of (\ref{consistency0}), as well as any non-linear effects and noise. To make progress however, we assume that the ``mode-coupling'' term, $\s_{MC}$, is Gaussian. This tells us that the optimal $\s_z$ can be obtained through Wiener filtering:
\be
\s_{z}^\Lambda(\kk)=W_D(\Bbbk)\bm{d}(\kk)\ ,\ \ \hbox{with}\ W_D(\Bbbk)=\frac{\langle \bm{s}_z^*(\kk)\cdot\bm{d}(\kk)\rangle}{\langle |\bm{d}(\kk)|^2\rangle}\ .
\ee
We put a superscript $\Lambda$ to highlight that the above equation recovers only the low-$k$ modes. Note that the optimal $W_D$ depends on the number of iterations $N$ and $M$, and as in the case of $W_\delta$ can be cheaply extracted from simulations. Using equation (\ref{ZA}), the low-$k$ linear density field is given by
\be\label{dL}
\delta_\Lambda^{\mathrm{new}}(\kk)=-i\s_{z}^\Lambda(\kk)\cdot\kk= -i W_D(\Bbbk)\bm{d}(\kk)\cdot \kk\ ,
\ee
where ``new'' denotes that the linear density reconstructed with the scheme proposed in this paper.  Note that the resulting large-scale reconstructed linear density field, $\delta_\Lambda^{\mathrm{new}}$, is consistently derived in Lagrangian space. This is in contrast to standard reconstruction, where the large-scale modes are reconstructed using a Zel'dovich-like approximation (\ref{sR}, \ref{deltaLambda}), which mixes Lagrangian and Eulerian coordinates.

We plot the filter $W_D$ (see next Section for description of the N-body simulations we use) in Fig.~\ref{fig:Wiener} for $z=0$ and 1, resulting after applying the scheme above with $N=M=5$. Similar to $W_\delta$, $W_D$ decays roughly on the non-linear scale, although the decay is steeper than for $W_\delta$. The filter features strong wiggles corresponding to the BAO wiggles. In order not to introduce those wiggles by hand into the reconstruction scheme, we remove them by fitting a Gaussian to $W_D$, and using that fit instead of $W_D$ in our calculations (e.g. in eq.~(\ref{dL})). We show the Gaussian fit to $W_D$ in Fig.~\ref{fig:Wiener} for reference (the curves labeled $W_{D,\mathrm{exp}}$). The Gaussian scale in $W_{D,\mathrm{exp}}$ is $2.80\,$Mpc$/h$ for $z=0$ and $1.54\,$Mpc$/h$ for $z=1$.

\subsubsection{Recovering the short-scale modes}

The procedure which we used above to obtain $\s_z^\Lambda$ is definitely non-optimal (To remind the reader the  pseudo-optimal scheme is given in eq.~(\ref{iter})). Thus, the residual $\bm{d}-\s_z^\Lambda$ contains important information about $\s_z$. There are many choices of how one can recover the short-scale modes. We find that a method that works well is to move a uniform field of particles (with positions $\q_u$) with the short-scale residual displacements, which in Fourier space are given by $\bm{d}_s(\kk)=(1-W_D(\Bbbk))\bm{d}(\kk)$. Thus, we have:
\be
\delta_s^{\mathrm{new}}=\delta[\q_u+\bm{d}_s(\q_u)]\ ,
\ee
which corresponds to the density field obtained by pulling back the data particles by the large-scale $\s_z^\Lambda$ as done in standard reconstruction. 
Note that all of the relevant information has been transformed (although arguably this is a lossy transformation) from $\delta$ to $\bm{d}$, and therefore we do not need to add the residual $\delta_R^{(N)}$ from (\ref{dR}). Indeed adding it, only degrades the performance of our method, while increasing $N$ makes $\delta_R^{(N)}$  from App.~\ref{app:iter} increasingly irrelevant.

\subsubsection{Putting it together}

The final linear density approximator for this first stage of the reconstruction is then given by
\be\label{finalS0}
\delta_L^{\mathrm{new}}(\q)=\delta_\Lambda^{\mathrm{new}}(\q)+\delta_s^{\mathrm{new}}(\q)\ .
\ee
Combining the above equation with (\ref{ZA}), we can find the displacement field $\s_z^{\mathrm{new}}$ corresponding to $\delta_L^{\mathrm{new}}(\q)$:
\be\label{sznew}
\s_z^{\mathrm{new}}=i\frac{\kk}{\Bbbk^2}\delta_L^{\mathrm{new}}\ .
\ee
 It is this $\s_z^{\mathrm{new}}$, which we then use as an initial guess for the pseudo-optimal second stage of the reconstruction, given by (\ref{iter}).

\section{Numerical results}\label{sec:results}

To gauge the performance of our reconstruction scheme against standard reconstruction, we ran 26 N-body simulations using the GADGET-2 code \cite{gadget}, with $256^3$ particles each, and a box size of $500\,$Mpc$/h$. The cosmological parameters  we use are (in the standard notation) as follows
\be\label{cosmo}
(\Omega_b,\Omega_{\mathrm{matter}},\Omega_\Lambda,h,n_s,\sigma_8)=(0.046,0.28,0.72,0.70,0.96,0.82)\ .
\ee
The initial conditions are set at a redshift of $49$, using the 2LPT code provided by \cite{2006MNRAS.373..369C}. 

Below we show results from the two-stage reconstruction scheme presented in this paper. The results for the first stage of the reconstruction are denoted by ``IRec'' in the figures that follow. That corresponds to the reconstructed linear density field from Section~\ref{sec:IC} (given by eq.~(\ref{finalS0})). That reconstructed linear density field is used to construct a guess displacement field $\s_z^{\mathrm{new}}$, (\ref{sznew}), which is then fed into the iterative solution for $\s_z$ given by (\ref{iter}) and then into (\ref{finalFix}). The final reconstructed Zel'dovich displacement field, $\hat \s_z$, obtained from (\ref{finalFix}) is converted to a reconstructed linear density field, again using (\ref{ZA}). That result is denoted by ``Rec'' in the figures that follow. We refer the reader to Appendix~\ref{sec:details} for further details on our numerical implementation of (\ref{iter}).

\begin{figure}
  \centering
  $z=0$\hspace{0.5\textwidth} $z=1$
  \\
  \subfloat{\includegraphics[width=0.5\textwidth]{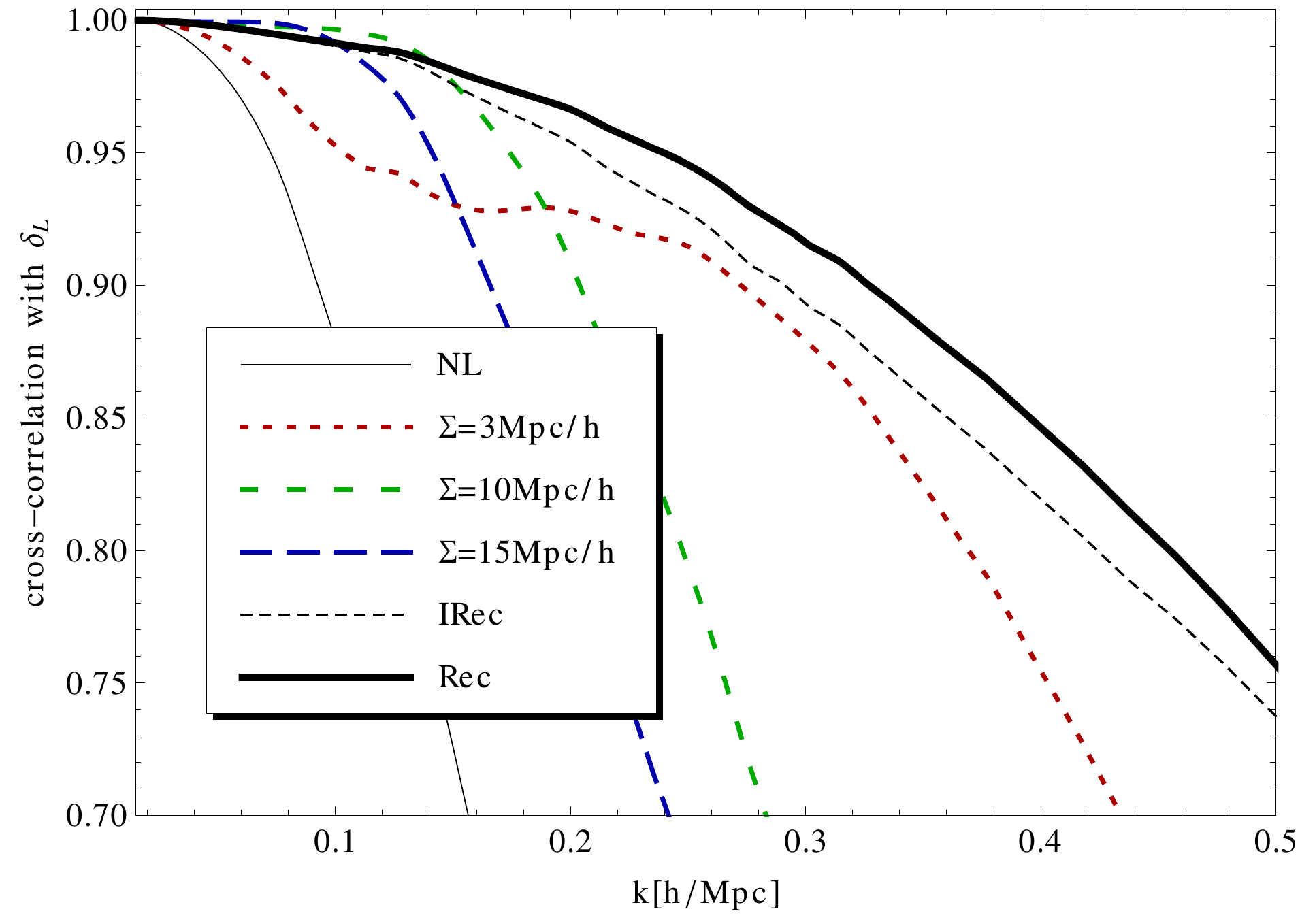}}                
  \subfloat{\includegraphics[width=0.5\textwidth]{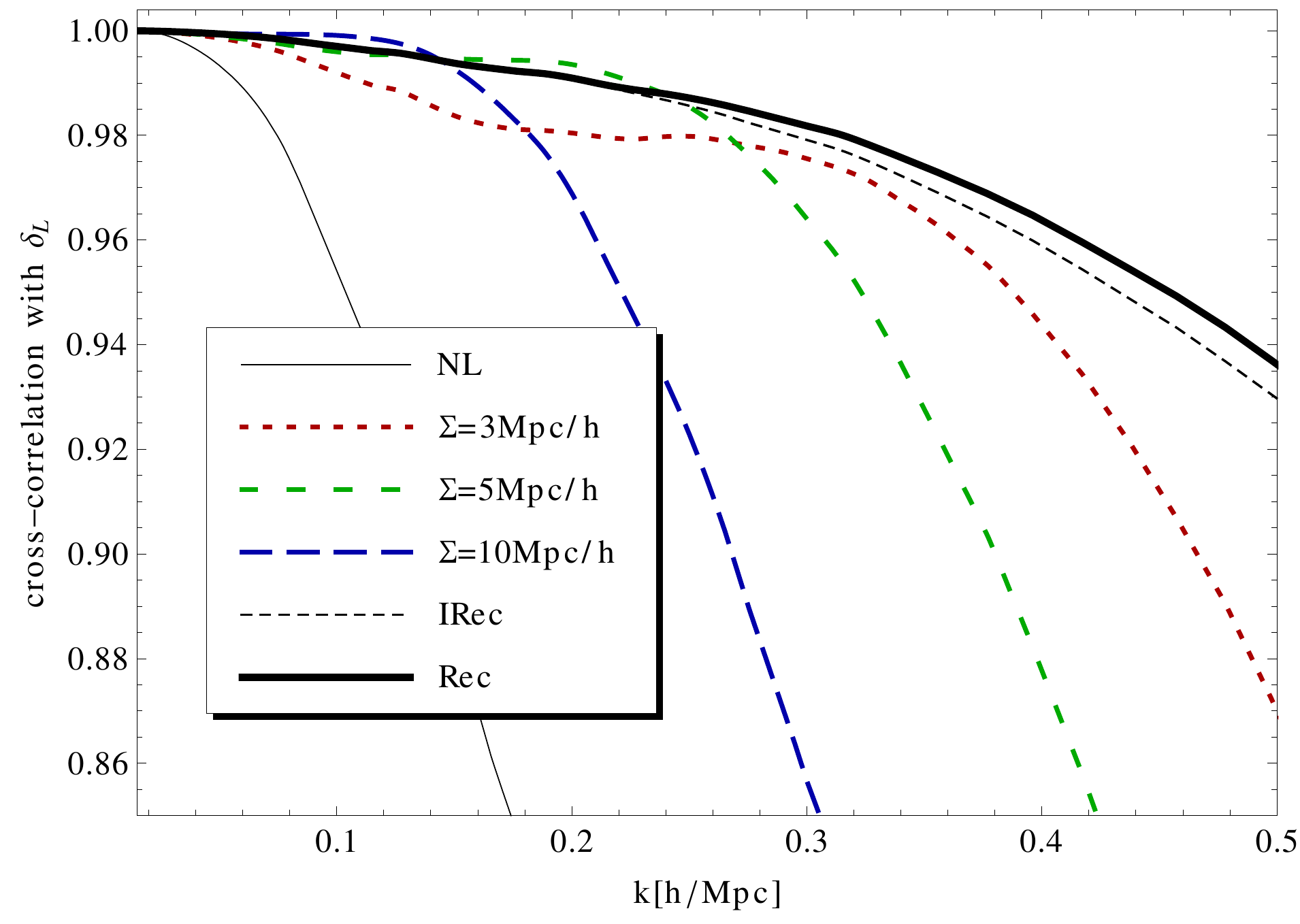}}
  \caption{Cross-correlation function between the linear density field and the density field labeled in the legend. The ``NL'' line corresponds to the cross-correlation coefficient between the linear and non-linear densities; while the ``Rec'' curve corresponds to the reconstruction scheme proposed in this paper. The curves labeled with different values of $\Sigma$, correspond to the standard reconstruction result with a Gaussian smoothing on the scale $\Sigma$. }
  \label{fig:cross}
\end{figure}

In Fig.~\ref{fig:cross} we show the cross-correlation between the linear, $\delta_L$, and the non-linear density $\delta$ (curve denoted by ``NL''), as well as the one between the linear and  reconstructed densities. The curves for the standard reconstruction result are denoted by the value of the smoothing length $\Sigma$ entering in $W_G$. The cross-correlation between our reconstructed linear density fields and the true linear density field is shown with the curves denoted by ``IRec'' and ``Rec''. 

One can see that standard reconstruction provides markedly different results for different $\Sigma$. In the limit of very large $\Sigma$, $W_G\to 0$ and we obtain $\delta_\Lambda=0$ and $\delta_s=\delta$. This implies  that the cross correlation between $\delta^{\mathrm{st}}_L$ and $\delta_L$ approaches the one between the linear and non-linear density fields -- a behavior which one can clearly see in Fig.~\ref{fig:cross}. In the limit of very small $\Sigma$, we obtain $\delta_s=0$ and a $\delta_\Lambda$ which according to the discussion at the end of Section~\ref{sec:st} can be roughly approximated by the density in the ZA. Therefore, the cross correlation between $\delta^{\mathrm{st}}_L$ and $\delta_L$ can be approximated by the one between the linear and ZA density, which is in turn well approximated by  the cross-correlation between the linear and non-linear density fields (e.g. \cite{tassev}). This behavior can be seen in the appearance of the step at low $k$ observed for $\Sigma=3\,$Mpc$/h$ in Fig.~\ref{fig:cross}. Therefore, by varying $\Sigma$ we can trade between reconstructing either the large-scale modes ($k\lesssim 0.2h/$Mpc for $z=0$); or the small-scale modes ($k\gtrsim 0.2h/$Mpc for $z=0$) at the expense of the large-scale modes.

Our reconstruction scheme does not treat the large-scale modes in the Zeldovich-like approximation of eq.~(\ref{deltaLambda}), and instead tries to reconstruct the true linear density through eq.~(\ref{ZA}). Thus, from Fig.~\ref{fig:cross} one can see that both stages of our reconstruction scheme recover both the large-scale modes along with the short-scale modes with high fidelity as measured by the cross-correlation. The first stage of our reconstruction is analogous to the standard reconstruction scheme, apart from our treatment of the large-scale modes. This implies that the main gain of our proposed methods comes from the fact that we obtain the large-scale linear power by applying eq.~(\ref{ZA}), i.e. by explicitly avoiding the mixing up of Eulerian and Lagrangian coordinates done in the standard scheme.

Note that the high-$\Sigma$ curves have slightly better cross-correlation at low $k$ than our reconstruction scheme. However, this does not help the reconstruction of the acoustic peak (as we will see below), because it is at the expense of a much poorer reconstruction of the small-scale modes.

\begin{figure}
  \centering
  $z=0$\hspace{0.5\textwidth} $z=1$
  \\
  \subfloat{\includegraphics[width=0.5\textwidth]{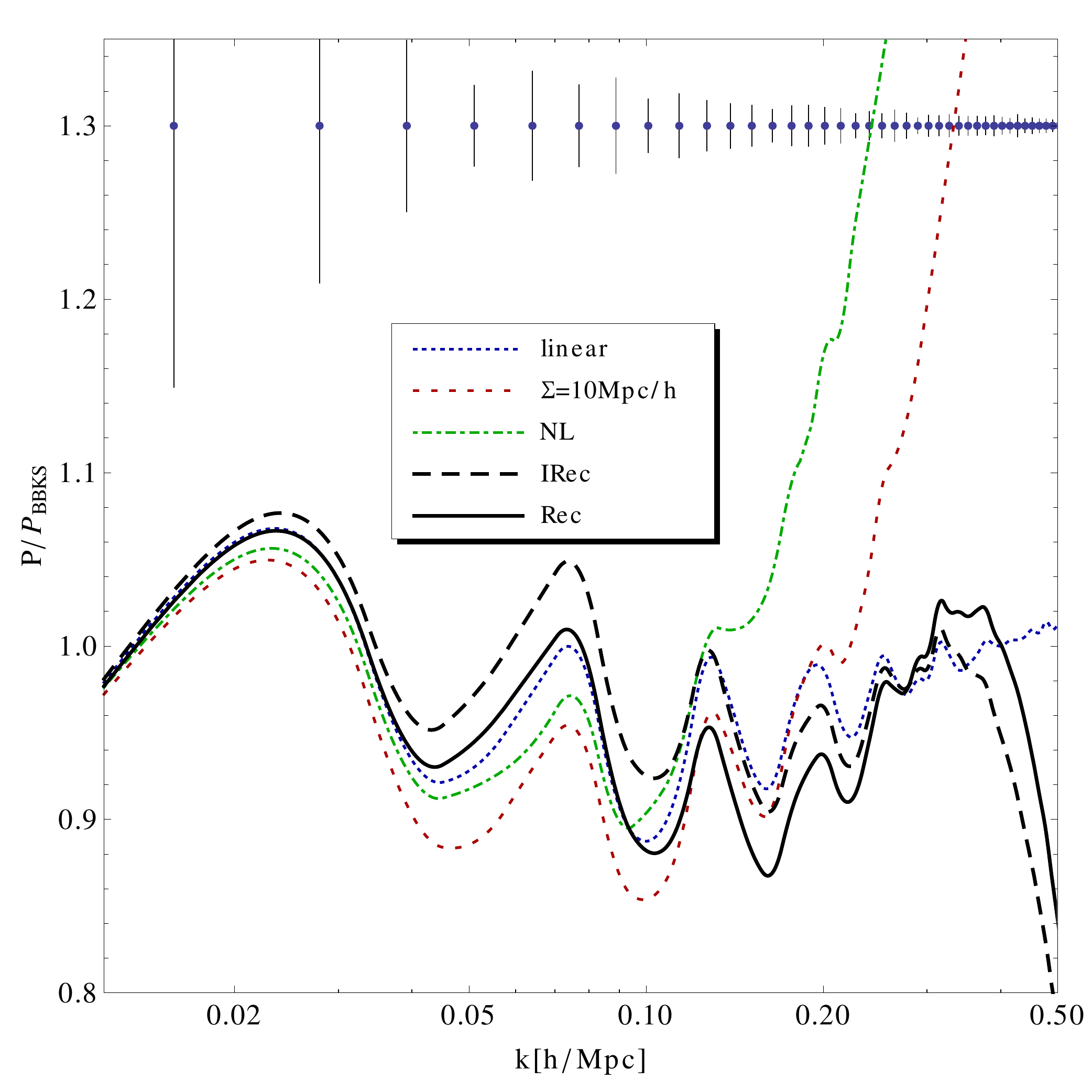}}                
  \subfloat{\includegraphics[width=0.5\textwidth]{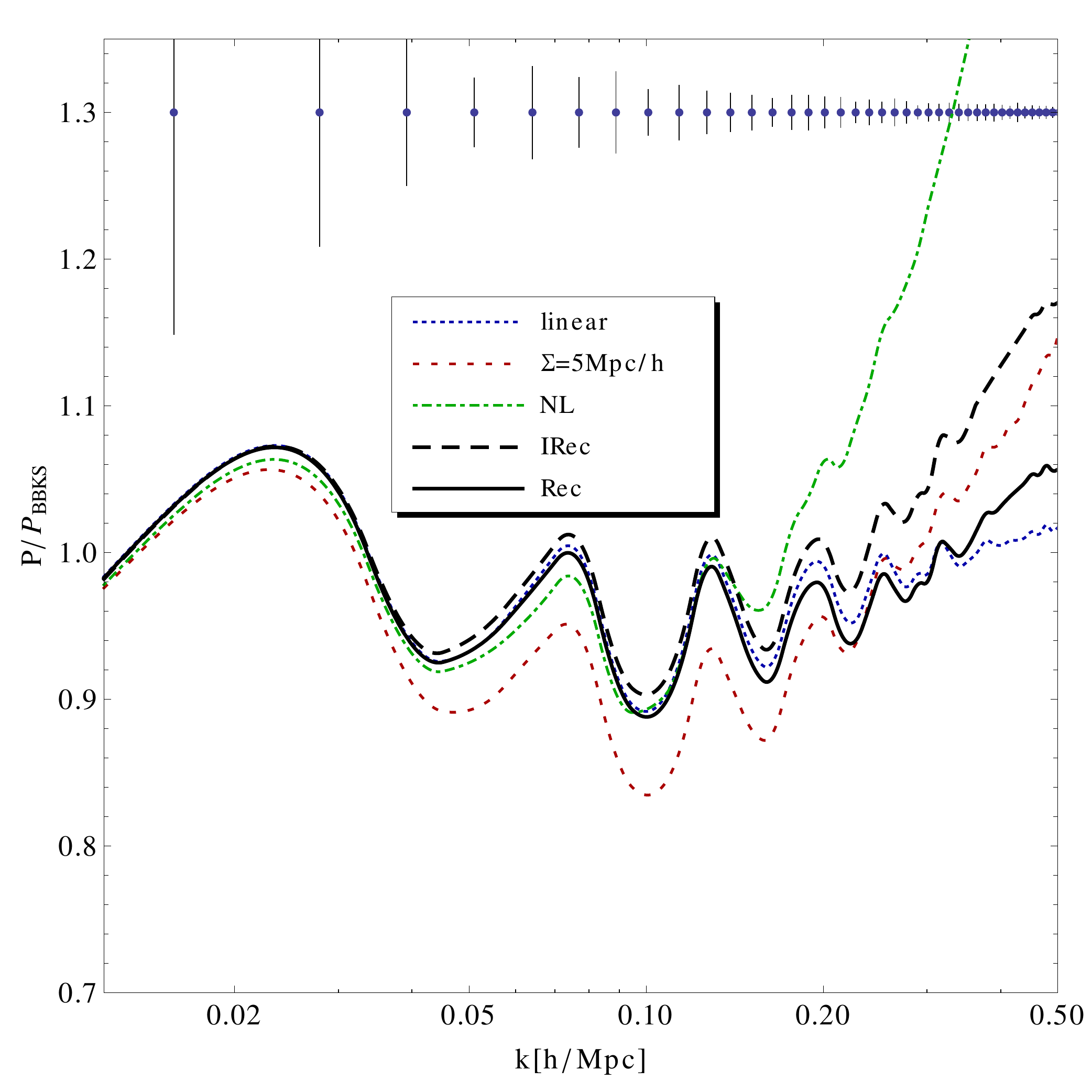}}
  \caption{Shown are the linear and non-linear (obtained from N-body simulations) density power spectra, as well as the power spectra of the reconstructed density fields. The standard reconstruction result is given by the curve denoted by $\Sigma$; while the result of this paper is denoted by ``Rec''. The errorbars show the 2-sigma errors in the average of the linear power-spectrum; and are representative of the errors in the rest of the power-spectra. The $k$-positions of the errorbars indicate the binning of the power spectra, while the curves were obtained after splining the result. Note that ideally the reconstructed density power spectrum should match the linear power, i.e. should not be simply within the errorbars. In practice, however, the BAO peak is determined by the period of the oscillations and is not affected by the broader-band features. }
  \label{fig:P}
\end{figure}

\begin{figure}
  \centering
  $z=0$\hspace{0.5\textwidth} $z=1$
  \\
  \subfloat{\includegraphics[width=0.5\textwidth]{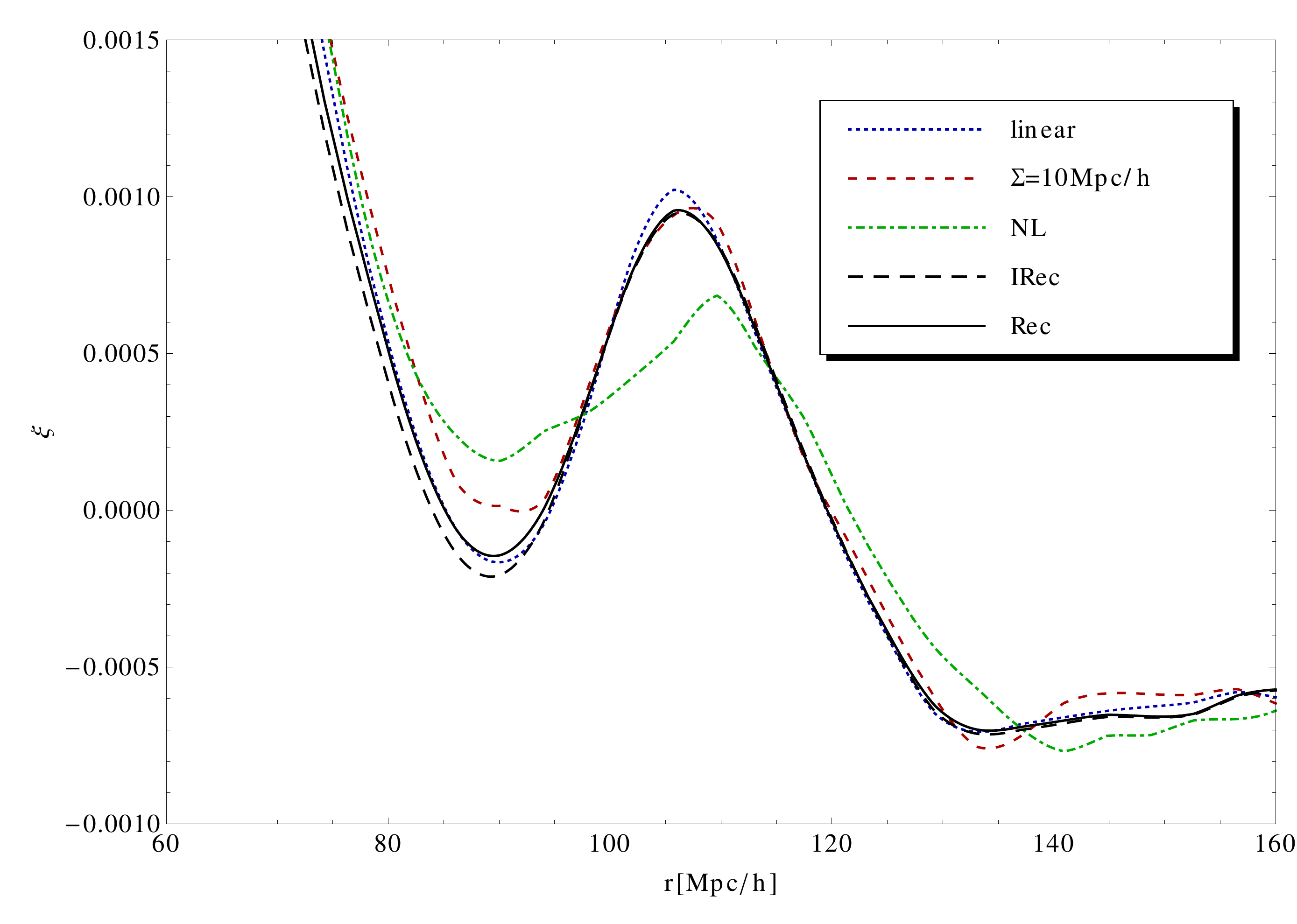}}                
  \subfloat{\includegraphics[width=0.5\textwidth]{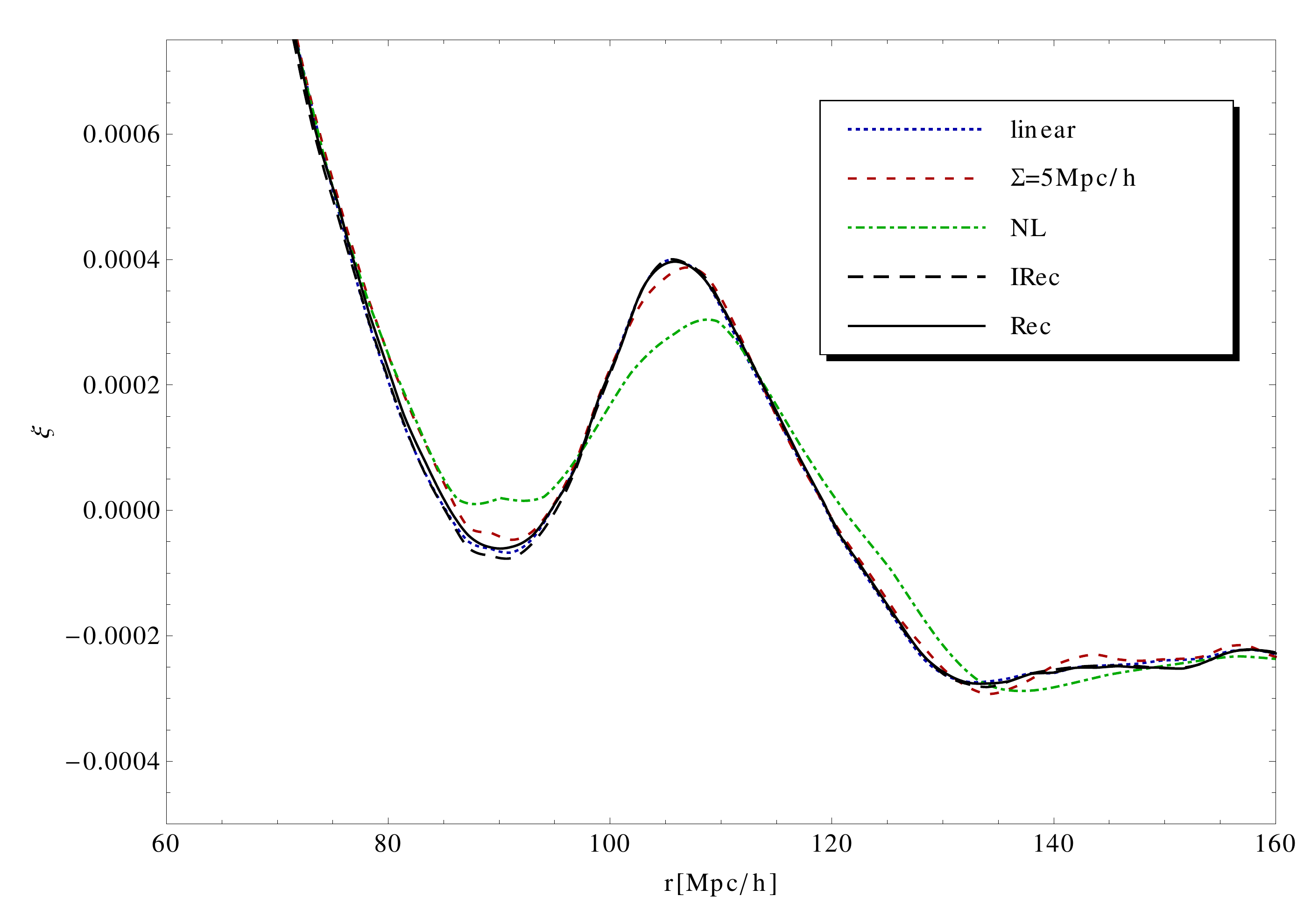}}
  \caption{Shown are the 2-pt matter correlation functions corresponding to the power spectra in Fig.~\ref{fig:P}. 
   Due to the missing large-scale power due to our relatively small box sizes, the 2-pt functions are shifted down relative to the true 2-pt functions, and a slight extra slope is introduced. However, since we compare results obtained from the same N-body simulations, these large-scale effects should not affect our conclusions. To reduce sample variance, the $z=0$ plot is obtained after binning the data in radial bins of width $3.9\,$Mpc$/h$; while the bins for $z=1$ are of width $1.95\,$Mpc$/h$ (apart from the NL curve which is binned in twice larger bins).} 
  \label{fig:xi}
\end{figure}

In Fig.~\ref{fig:P} we show the linear and non-linear density power spectra, as well as the power spectra of the reconstructed density fields. The corresponding 2-point functions are shown in Fig.~\ref{fig:xi}. From Fig.~\ref{fig:cross} we chose $\Sigma=10\,$Mpc$/h$ for $z=0$, and $\Sigma=5\,$Mpc$/h$ for $z=1$ as fiducial values for the smoothing length in the standard reconstruction scheme. Note that the reconstructed 2-pt function  is rather insensitive to the choice of $\Sigma$ -- the fiducial $\Sigma$ values corresponding approximately to the best-reconstructed acoustic peak with that scheme.

Ideally, the reconstructed density power spectrum should match the linear power spectrum. The standard reconstruction scheme  recovers the BAO wiggles up to\footnote{This maximum $k$ for which we see wiggles in the power spectrum depends on the smoothing scale $\Sigma$. One can reduce $\Sigma$ by about a factor of 2 and recover wiggles with standard reconstruction up to $k\sim 0.35h/$Mpc for $z=0$. However, that is at the expense of reducing the power for scales $k\sim0.1h/$Mpc, due to the fact that standard reconstruction treats the large-scale modes in a Zel'dovich-like approximation. This results in an acoustic peak, which is more smeared than for our fiducial choice of $\Sigma$.} $k\approx k_{NL}=0.25h/$Mpc at $z=0$. The reconstruction scheme proposed in this paper recovers the baryon wiggles up to $k\approx0.4h/$Mpc at $z=0$ after the pseudo-optimal second stage of the reconstruction (and up to $k\approx0.35h/$Mpc after the first stage), which corresponds to roughly two more well-reconstructed wiggles in the power spectrum.\footnote{One might worry that the first stage of our reconstruction gives better match to the linear power spectrum at $k\approx 0.15h/$Mpc than our second stage. However, this seems accidental, since if one ``fixes'' the large-scale power in the first-stage result (e.g. by multiplying by a broad-band Gaussian decay), this result would go away.} Therefore, the peak in the corresponding 2-pt function (see Fig.~\ref{fig:xi}) shows marked improvement compared with the result from standard reconstruction. 

The improvements  for $z=1$ in the 2-pt function and the power spectrum with our scheme are not that stark when compared to the results from the standard reconstruction scheme. Still, one should note that  the power spectrum that we recover for $z=1$ lies very close (within a couple of percent for our choice of binning) to the true linear power up to $k\approx 0.4h/$Mpc (after the second stage of our reconstruction scheme), testifying to the quality of the reconstruction scheme proposed in this paper.

\section{Summary}\label{sec:summary}

In this paper we derive a pseudo-optimal scheme for reconstructing the baryon oscillations in the matter power spectrum. The proposed scheme has no free parameters (apart from the grid size used for the Cloud-in-Cell assignments). It is based on a simple model (see (\ref{master})) for the non-linear density proposed in \cite{2012JCAP...12..011T}.  

The reconstruction scheme of this paper is made up of two stages. The first consists in finding a good guess for the Zel'dovich displacement field. That field is then fed to the second stage, which solves for the Zel'dovich displacement field that maximizes the Likelihood function for our model for the non-linear density field. The first stage requires $\mathcal{O}(10)$ iterations, while the second stage -- $\mathcal{O}(3)$.

The main differences between the reconstruction scheme proposed here and the standard reconstruction method is 1) in using optimal filters in both stages of the reconstruction; and 2) in our treatment of the large-scale modes. The first point implies that our method has no free parameters such as the smoothing scale entering in the standard reconstruction scheme. Regarding the second point, standard reconstruction obtains a large-scale density field which is in fact evaluated using a Zel'dovich-like approximation. That choice degrades the reconstruction of the large scales especially for the low values of the filtering scale which are needed if one wants to recover the short scales better. Such a trade-off is in practice absent in our method, which also avoids the mixing up of Lagrangian and Eulerian space coordinates, which is done in the standard scheme.

We find that for $z=0$ the reconstruction scheme proposed in this paper recovers the baryon wiggles up to $k\approx0.4h/$Mpc, compared with $k\approx0.25h/$Mpc for the standard reconstruction scheme. This implies that we recover roughly two more wiggles in the power spectrum, which results in a markedly improved sharpening in the reconstructed acoustic peak compared with the result from standard reconstruction. However, we postpone a proper error analysis for the reconstructed acoustic peak position to future work.

 We find that even the first stage of our  scheme yields excellent reconstruction, recovering the baryon wiggles up to $k\approx0.35h/$Mpc for $z=0$.
Whether this result will persist for biased tracers, remains to be seen. 
%However, we argue that for biased tracers, the second stage of our reconstruction algorithm may prove far more important, since it does not assume that those biased tracers sample Lagrangian space uniformly.

One can envision numerous possible improvements of the scheme proposed in this paper. One such improvement is using the model in \cite{2012JCAP...12..011T} for the non-linear density field based on second-order Lagrangian perturbation theory, instead of the one used in this work based on the ZA. We postpone such analysis, as well as the application of the proposed method to biased tracers in redshift space, to future work.

\appendix

\section{Implementing (\ref{iter}) numerically}\label{sec:details}

To use (\ref{iter}), we need to map $\x$ to $\q$ between the Fourier transforms $\mathcal{F}_{\kk\q}\mathcal{F}^{-1}_{\x\k}Q(\k)$, where $Q$ can be read off from (\ref{iter}). This is done by displacing a uniform grid of particles at $\q_u$ by  $(R_z*\s_z)(\q_u)$ (or by $(R_z*\s_z)(\q_u+\tilde\q)-(R_z*\s_z)(\tilde\q)$ for the Fisher matrix). Then we interpolate the quantity $\mathcal{F}^{-1}_{\x\k}Q(\k)$ to the particle positions using three-linear interpolation, and assign the obtained values to the initial $\q_u$. The resulting field corresponds to evaluating $\mathcal{F}^{-1}_{\x\k}Q(\k)$ in Lagrangian space and is therefore ready for feeding into $\mathcal{F}_{\kk\q}$. 

Another thing to note is that to make the algorithm stable we truncated $Q$ in the first derivative of $\L$ (the quantity in the inner square brackets in the first line of (\ref{iter})) at 1/2 the Nyquist wavevector, corresponding to $k\approx0.8h$/Mpc. This is reasonable since that quantity involves derivatives in Eulerian space. We checked the sensitivity of our method to this choice of cutoff, by pushing the cutoff to $1h/$Mpc. This resulted in only a minor change of the amplitude of the reconstructed $P_L$ at the high $k$ end, which did not affect the 2-pt or the cross-correlation functions.

Note that for all iterations of (\ref{iter}) we evaluate $W$ at the maximum of the likelihood function, using the result shown in Fig.~\ref{fig:Filters}, but with the wiggles removed. Given the excellent guess coming from the first stage of our reconstruction, this does not affect the final result, at the same time speeding up the calculation appreciably.

One final modification that we needed to do to (\ref{iter}) to make it more stable, was to multiply $W$ by a constant, call it $\gamma$. This is a standard trick used with the Newton-Raphson algorithm when applied to non-quadratic functions\footnote{In fact, the presence of $\gamma$ may be required even for quadratic functions to make the algorithm convergent, since we use the expectation value of the Hessian, and not the Hessian itself.}, such as our  $\log\L$. We find that  an initial $\gamma=0.2$ works well for our guess $\s_z$ coming from the first stage of the reconstruction, described in Section~\ref{sec:IC}. We restart an iteration with a $\gamma$ reduced by a half whenever the ratio of the rms displacement corresponding to $\s_z$ to the rms displacement corresponding to the correction (the term proportional to $W$ in (\ref{iter}) but before multiplying it by $\gamma$) is smaller than 0.95 times the ratio of the previous iteration. 

We find that this ratio of rms displacements is always bigger than one and quickly improves. We stop the iterations in (\ref{iter}) when that ratio exceeds 10 for $z=0$ and $30$ for $z=1$, since we find that further iterations do not improve the result further. This stop condition is reached after $\mathcal{O}(3)$ iterations, given our initial guess for $\s_z$.

\section{Solving eq.~(\ref{consistency0}) for the displacement field starting with eq.~(\ref{consistency}).}\label{app:iter}

In this section, we solve eq.~(\ref{consistency0}) for $\bm{d}(\q_p)$ starting with eq.~(\ref{consistency}).
First, we construct a displacement field on a grid, which is initially set to zero: $\g^{(0)}=0$. We then apply the following iterative scheme on $\g$ in a way analogous to the method described in \cite{2010ApJ...720.1650S}: 
\be
\g^{(n)}(\k)=i\frac{\k}{k^2}W_\delta(k) \delta_R^{(n-1)}(\k)\ ,
\ee
where 
\be\label{dR}
 \delta_R^{(n)}=\delta[\x_p^{(n)}]\ , \ \ 
\x_p^{(n)}=\x_p^{(n-1)}-\g^{(n)}(\x_p^{(n-1)})\ .
\ee
In writing down the equations above we applied the Wiener filter, $W_\delta$, entering in eq.~(\ref{delta1}). We evaluate $\g^{(n)}$ at $\x_p^{(n-1)}$ by using three-linear interpolation.
The initialization is given by $\x_p^{(0)}=\x_p$, i.e. we start with the true particle positions, which implies that $\delta_R^{(0)}=\delta$, and that $\bm{g}^{(1)}(\k)$ approximately matches $\bm{d}$ in (\ref{consistency}).

As discussed before (after eq.~(\ref{delta1})), each successive iteration in the above algorithm kills the large-scale power in $\delta_R$ to higher and higher $k$. So, after 5 iterations, the power in $\delta_R$ for $z=0$ is only about 10\% of the power in $\delta_L$ at $k\sim 0.75h/$Mpc and is strongly suppressed at lower $k$, which implies that for the purposes of the acoustic peak reconstruction, we satisfy eq.~(\ref{consistency0}).

The total displacement by which we have moved particle $p$ after $N$ iterations is given by:
\be
\SS_p^{(N)}=\sum_{n=1}^N \g^{(n)}(\x_p^{(n-1)})\ ,
\ee
which is a vector which should be anchored to the particle position after the $N$-th iteration, $\x_p^{(N)}$.

\section{Interpolating the displacements to a grid}\label{app:interp}

We would like to construct an approximation for $\bm{d}(\q)$ starting from the pairs $(\x_p^{(N)},\SS_p^{(N)})$ which approximately equal  $(\q_p,\bm{d}(\q_p))$ up to some short-scale corrections.  One can do that in a variety of ways, e.g. using Voronoi tessellation to perform the interpolation to a grid. For the illustrative purposes of this paper, however, we choose a simple iterative algorithm: We use a weighted CiC assignment to construct the mass-weighted displacement field on a grid. Dividing that by $(1+\delta_R^{(N)})$ we obtain a guess displacement field on a grid. Call that $\G^{(1)}(\q)$.
Then we perform the following iterative procedure:
\be
\SS_p^{(N,m)}=\SS_p^{(N,m-1)}-\G^{(m)}(\x_p^{(N)})\ , \ \ \mathrm{with}\ \  \SS_p^{(N,0)}\equiv \SS_p^{(N)}\ ,
\ee
where $\G^{(m)}$ is obtained by dividing the mass-weighted displacement field corresponding to the pairs $(\x_p^{(N)},\SS_p^{(N,m-1)})$ by the density $(1+\delta_R^{(N)})$. To avoid singularities, we set the displacement in zero density regions to zero. After $M$ iterations (in our case we find that $M=5$ works well enough), the approximate displacement field in Lagrangian space is given by
\be
\bm{d}(\q)=\sum_{m=1}^M \G^{(m)}(\q)\ .
\ee

\acknowledgments We would like to thank Uros Seljak, Nikhil Padmanabhan and Martin White for useful comments on the manuscript. The work of MZ is supported by NSF grants PHY-0855425, AST-0506556 \& AST-0907969, and by the David \& Lucile Packard and the John D. \& Catherine T. MacArthur Foundations.

\bibliography{mildly_NL_v1}

\end{document}